\documentclass{nature}

\usepackage{graphicx, amssymb,color,float}
\newfloat{raw}{h}{lop}
\usepackage[normalem]{ulem}
\pdfoutput=1

\title{Interfacial mode coupling as the origin of the enhancement of T$_{c}$ in FeSe films on SrTiO$_{3}$}

\author
{J. J. Lee$^{1,2,\dagger}$F. T. Schmitt$^{1,\dagger}$
R. G. Moore$^{1,\dagger}$
S. Johnston$^{3,4,5}$
Y.-T. Cui$^{1}$
W. Li$^{1}$\\
M. Yi$^{1,2}$
Z. K. Liu$^{1,2}$
M. Hashimoto$^{6}$
Y. Zhang$^{1,7}$
D. H. Lu$^{6}$\\
T. P. Devereaux$^{1}$
D. -H. Lee$^{8,9}$
Z.-X. Shen$^{1,2}$
\\
\footnotesize{$^1$Stanford Institute for Materials and Energy Sciences,}\\ 
\footnotesize{SLAC National Accelerator Laboratory, Menlo Park, CA 94025, USA}\\
\footnotesize{$^2$Departments of Physics and Applied Physics, and Geballe Laboratory for}\\
\footnotesize{Advanced Materials, Stanford University, Stanford, CA, 94305, USA}\\
\footnotesize{$^3$Department of Physics and Astronomy, University of British Columbia, Vancouver BC, Canada V6T 1Z1}\\
\footnotesize{$^4$Quantum Matter Institute, University of British Columbia, Vancouver BC, Canada V6T 1Z4}\\
\footnotesize{$^5$Department of Physics and Astronomy, University of Tennessee, Knoxville, Tennessee 37996-1200, USA}\\
\footnotesize{$^6$Stanford Synchrotron Radiation Lightsource, SLAC National Accelerator Laboratory, Menlo Park, CA 94025, USA}\\
\footnotesize{$^7$Advanced Light Source, Lawrence Berkeley National Lab,Berkeley, California 94720, USA}\\
\footnotesize{$^8$Department of Physics, University of California at Berkeley, Berkeley, CA 94720, USA}\\
\footnotesize{$^9$Material Science Division, Lawrence Berkeley National Laboratory, Berkeley, CA 94720, USA}\\
\\
\footnotesize{$^{\dagger}$ These authors contributed equally to this work.}\\
\normalsize{}
}

\date{}

\makeatletter
\newenvironment{EDfigure}{\let\caption\NAT@EDfigcaption}{}
\newcounter{EDFigureCounter}
\newcommand{\NAT@EDfigcaption}[2][]{{%
    \refstepcounter{EDFigureCounter}
    \ifthenelse{\value{EDFigureCounter}=1}{
        \newpage\noindent%
    }{
        \par\vfill
    }
    \sffamily\noindent\textbf{Extended Data Figure \arabic{EDFigureCounter}}\hspace{1em}#2}
    }
\makeatother
\begin{document}




\maketitle

\begin{abstract}
Single unit cell films of iron selenide (1UC FeSe) grown on SrTiO$_3$ (STO) substrates have recently shown superconducting energy gaps opening at temperatures close to the boiling point of liquid nitrogen (77 K)\cite{Wang_CPL12,Liu_NC12, He2013, Tan2013}, a record for iron-based superconductors. Towards understanding why Cooper pairs form at such high temperatures, a primary question to address is the role, if any, of the STO substrate.  Here, we report high resolution angle resolved photoemission spectroscopy (ARPES) results which reveal an unexpected and unique characteristic of the 1UC FeSe/STO system: shake-off bands suggesting the presence of bosonic modes, most likely oxygen optical phonons in STO\cite{PhysRevB.77.134111, STOPhonon_Ferroelectrics, INS_Stirling_PhononSTO}, which couple to the FeSe electrons with only small momentum transfer. Such coupling has the unusual benefit of helping superconductivity in most channels, including those mediated by spin fluctuations\cite{PhysRevB.54.14971, PhysRevB.89.134507, PhysRevB.69.094523, PhysRevB.82.064513, PhysRevB.54.R6877, PhysRevB.83.092505, Santi1996253}.  Our calculations suggest such coupling is responsible for raising the superconducting gap opening temperature in 1UC FeSe/STO. This discovery suggests a pathway to engineer high temperature superconductors.
\end{abstract}

\section{Main}
The dramatic enhancement of the superconducting transition temperature in FeSe from 8 K in bulk\cite{Hsu_PNAS08} to nearly 70 K when grown as a single layer on STO has generated tremendous interest as it raises the question of how Cooper pairing can be strengthened in non-bulk systems. We begin by studying the electronic structure of the 1 UC FeSe/STO film. Growth details are discussed in the Supplementary Information (SI) and Extended Data (ED) Fig. 1. The electronic structure is plotted in Fig. 1a-f and is consistent with previous reports\cite{Liu_NC12,Tan2013, He2013}.  The Fermi surface of the 1UC film consists of electron-like (concave up, labeled \textbf{A}) pockets centered around the Brillouin zone corner (M-point) with a band bottom 60 meV below the Fermi energy (E$_{F}$).  Further analysis reveals there are in fact two nearly-degenerate electron bands at M (see SI).  Below the bottom of \textbf{A},  one hole-like (concave down, labeled \textbf{B}) band is clearly resolved.  The zone center ($\Gamma$) consists of another hole-like band (labeled \textbf{D}) with a top located 80 meV below E$_F$.  The temperature evolution of the band structure is shown in Fig. 2a-f and shows typical superconducting gap behavior, with the electron band backbending at the Fermi momentum ($k_{F}$). Symmetrized energy distribution curves (EDCs) shown in Fig. 2c are fit using a phenomenological model (Ref. [\citen{PhysRevB.57.R11093}]) to reveal a 13 meV gap plotted in Fig. 2d.  Utilizing a mean-field (MF) formula we obtain  T$_c$ = 58$\pm$7 K, consistent with previous results within experimental uncertainty\cite{Liu_NC12,Tan2013, He2013}.  Analyzing the energy gap at different points (Fig. 2e-f) we observe a round, un-nested Fermi surface with a nearly uniform gap, making it unlikely that the gap is caused by other instabilities such as charge density waves.

We now turn to the most unexpected aspect of our 1UC data: the clearly resolved replica bands labeled \textbf{A$^\prime$}, \textbf{B$^\prime$} in Fig. 1d-1f. Aside from an approximately 100 meV relative shift in energy, all features of \textbf{A} are replicated in \textbf{A$^\prime$}. This includes the back-bending near $k_F$, seen in Fig. 3. Similarly \textbf{B} and \textbf{B$^\prime$} are also separated by the same energy shift with nearly the same dispersion.  In Fig. 1d we mark the peaks associated with \textbf{A}, \textbf{A$^\prime$} and \textbf{B}, \textbf{B$^\prime$} by blue and red, respectively, illustrating the clear presence of these features even in the raw EDC data. In the SI we rule out the possibility that such replicas can be caused by band structure effects, e.g. quantum well states\cite{Paggel_PRB00, Tang_PRL06}, and further discuss weaker features such as \textbf{C} and \textbf{D$^\prime$}. 

We believe the replica bands are due to the shaking off of quanta of the bosonic modes in STO, reminiscent of the vibron shake offs in the photoemission spectra of H$_2$ molecules\cite{TurnerHydrogen}. In the SI we make the case that such bosonic modes are associated with a high energy STO oxygen phonon band\cite{PhysRevB.77.134111, STOPhonon_Ferroelectrics} near 100 meV, although its exact energy may be modified somewhat by the presence of the overlaying film.  This identification is supported by recent ARPES on STO surface states which show a phonon-induced hump at approximately 100 meV away from the main band\cite{PhysRevB.81.235109} and through inelastic neutron scattering\cite{INS_Stirling_PhononSTO}. Resolving replicas of an entire band with such a clear dispersion, as seen in Fig. 1d, is unprecedented in a solid. This becomes possible here by the substantial electron-phonon (e-ph) coupling, the fact that the collective mode energy is greater than the width of the electron band below E$_F$, and most importantly, the fact that the electron-phonon coupling only allows small momentum transfer to the electron (see later discussion and SI).  

We now turn to the multi-UC films. We have measured 2UC and 30UC thick FeSe films grown on similar Nb-doped STO substrates. In Fig. 3a-f we compare the second energy derivative  of the APRES spectra for the 1UC (a,d) , 2UC (b,e) and 30UC (c,f) films. We find that the 2UC and 30UC band structure consists of both electron-like and hole-like bands crossing E$_F$ near M, similar to what has been observed for bulk FeSe\cite{Maletz2013}. This is dramatically different from the 1UC film's band structure, where only electron bands cross E$_F$. This implies that the 1UC film is much more heavily electron doped than even the 2UC film\cite{Tan2013}. Most importantly we observe neither a superconducting-like energy gap nor replica bands in the multi-UC films. A complete comparison of the band structures can be found in ED Fig. 2.  The fact that we find only the 1UC film has the superconducting gap is consistent with the conclusion of Ref. [\citen{Wang_CPL12}], where despite transport data showing multi-UC films to be superconducting, it is argued that only the 1UC portion of the film actually superconducts.

The temperature evolution of the replica bands in the 1UC film is plotted in Fig. 3g-j (additional temperatures are plotted in ED Fig. 3). We see that these replica bands, and hence this anomalous electron-phonon coupling, persists to temperatures significantly above the gap-opening temperature. The contrast between the 1UC and multiple-UC films suggests that the STO phonon which causes the replica bands is also responsible for enhancing Cooper pairing\cite{PhysRevB.86.134508}. The fact that the replica bands in the 1UC film follows the dispersion of the main band so closely suggests that, upon either absorption or emission, the phonons can only transmit small momenta to the electron. Such a strong forwardly-focused e-ph interaction is unusual, as it can enhance Cooper pairing in most symmetry channels, including those with a sign change (see SI and ED Fig. 6)\cite{PhysRevB.89.134507, PhysRevB.82.064513, PhysRevB.54.14971, PhysRevB.69.094523,PhysRevB.54.R6877, PhysRevB.83.092505, Santi1996253}.  

To estimate the strength of the e-ph coupling, we perform a high statistics scan at M at low temperature, plotted in Fig. 4a. Using a spline background, we find a lower bound of 1/6 for the intensity ratio of the replica band to the main band (see Fig. 4b and ED Fig. 4). We then take the intensity ratio as input and theoretically estimate the e-ph coupling strength (see SI). Plotted in Fig. 4c is a simulated spectral function calculated using a model where both the electron and hole bands couple to a flat phonon band with energy 80 meV. By tuning the coupling strength and the forward-focusing parameter we can well-reproduce a band-replica separation of approximately 100 meV. The simulated EDC is plotted in Fig. 4b. The ARPES spectrum is well-reproduced, especially the abrupt loss in spectral weight of band \textbf{A$^{\prime}$} beyond a certain momentum window, and the momentum broadening of the bands, plotted in ED Fig. 5.


From the intensity ratios we obtain the e-ph coupling constant $\lambda\approx0.5$, which is substantial considering only a narrow range of phonon modes at such high frequency contribute to the coupling (due to nearly forward scattering). This estimate in turn yields an effective phonon-mediated attraction strength $v_{\mathrm{eff}} \approx$ 10 meV (see SI). Under the assumption that Cooper pairing is caused by the magnetic interaction in the absence of the e-ph interaction, and following the effective Hamiltonian approach described in Ref. [\citen{Davis29102013}], we first determine the pairing symmetry without the e-ph interaction as a function of the ratio between the nearest and second neighbor magnetic exchange constant $J_1$ and $J_2$. The band structure we use is plotted in ED Fig. 7\cite{PhysRevB.86.134508}. It is noted that for the physically relevant $J_2$/$J_1$ ratio (\textgreater0.5) the pairing is in-phase $s$-wave between the two electron pockets \cite{PhysRevB.88.100504, NatComms.4.2783}, see ED Fig. 8. We then determine the enhancement of the Cooper pairing temperature as a function of the ratio between $v_{\mathrm{eff}}$ and the antiferromagnetic exchange constant $J = \sqrt{(J_1^2 + J_2^2)}$ (see SI). We obtain $J\sim30$ meV from the largest nearest- and second-neighbor exchange constants determined from neutron scattering experiments known to us\cite{NeutronBaKFeAs}, thus overestimating the magnetic coupling. Fig. 4d plots the T$_c$ enhancement as a function of $v_{\mathrm{eff}}/J$ for varying ratios of \textit{J}$_2$ to \textit{J}$_1$. Using the $v_{\mathrm{eff}}$ extracted from our data and the above estimate of J, we determine the enhancement factor to be $\approx$ 1.5. This enhancement is a lower bound as we use the most conservative estimate of the e-ph coupling, and the largest $J$ value. For bulk materials with similar band structures, e.g. K$_x$Fe$_{2-y}$Se$_2$\cite{Zhang2011}, or materials from which we obtain the above stated $J$ value\cite{NeutronBaKFeAs}, the T$_c$'s range from 30-40 K.  Multiplying by the enhancement factor yields a gap-opening temperature in fairly good agreement with our films. It should be noted that we do not use the T$_c$ of bulk FeSe due to the fact that our films are heavily electron doped and hence have a very different band structure from that of bulk FeSe.

Motivated by the physical picture presented here, we propose a heterostructure where 1UC FeSe is sandwiched between STO on both sides, effectively doubling $v_{\mathrm{eff}}$. A simple reading of Fig. 4d suggests a T$_c$ enhancement of $\sim$2.5, placing the Cooper pairing temperature well above the liquid nitrogen temperature. 

\section{References}
\bibliographystyle{naturemag}

Supplementary Information is linked to the online version of the paper at www.nature.com/nature.

\textbf{Acknowledgments} This work was supported by the U.S. Department of Energy, Office of Science, Basic Energy Sciences, Materials Sciences and Engineering Division. D.-H. Lee is supported by DOE Office of Basic Energy Sciences, Division of Materials Science, under Quantum Material program, DE-AC02-05CH11231. Measurements were performed at the Stanford Synchrotron Radiation Lightsource, a national user facility operated by Stanford University on behalf of the U.S. Department of Energy, Office of Basic Energy Sciences.

\textbf{Author Contributions} J.J.L., F.T.S. and R.G.M. grew films, collected and analyzed data, and wrote the paper. S.J. and D.H.Lee performed theory calculations. Y.T.C, W.L., Z.K.L., Y.Z., D.H.Lu and M.Y. provided discussion about data and interpretation. M.H. and D.H.Lu provided experimental support at SSRL. Project direction was provided by D.H.Lee, T.P.D, and Z.X.S.

\textbf{Author Information} Reprints and permissions information is available at www.nature.com/reprints. The authors declare no competing financial interests. Correspondence and requests for materials should be addressed to Z.X.S. at zxshen@stanford.edu.

\begin{figure}
\includegraphics[scale=1]{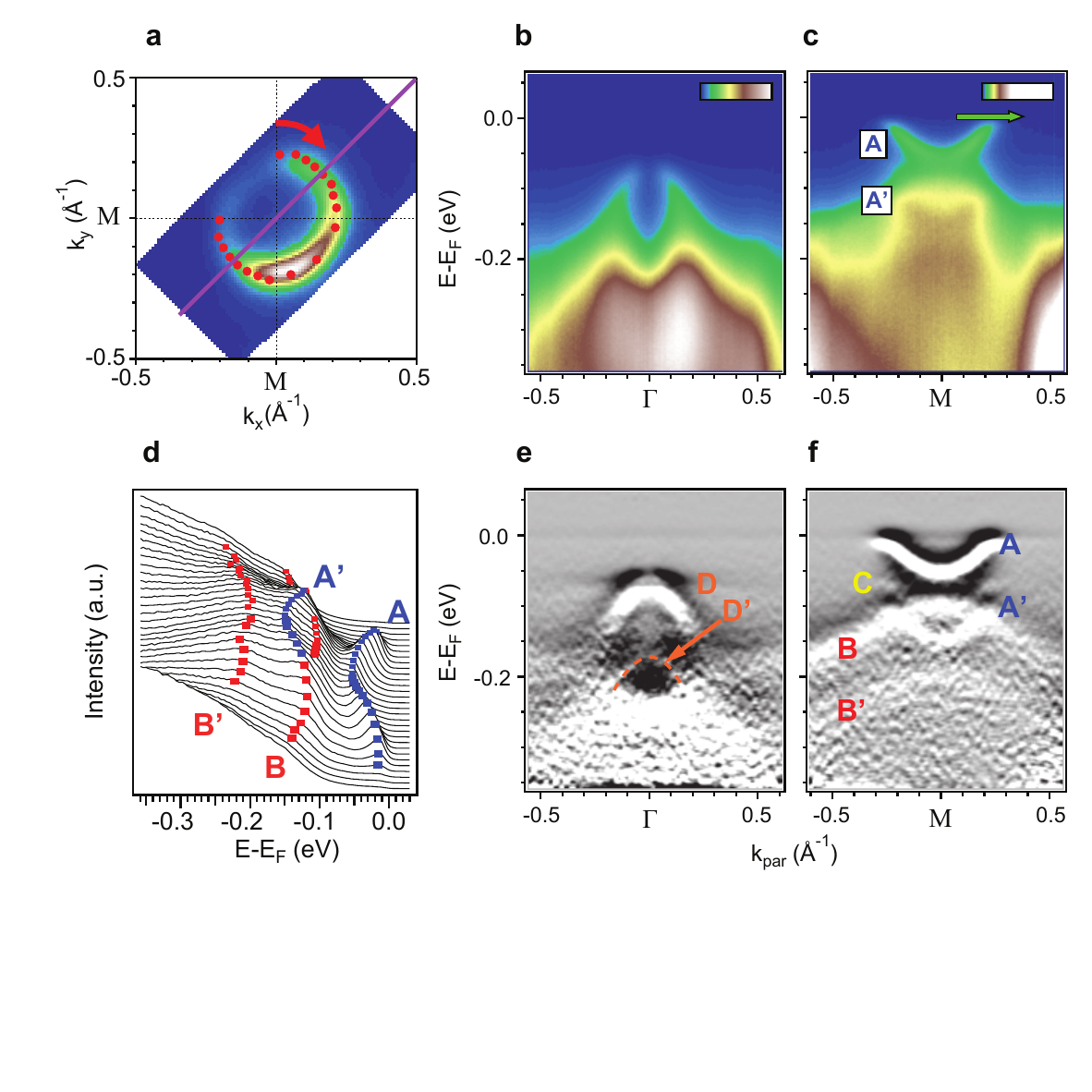}
\caption{\textbf{Fermi surface map and high symmetry cuts of 1UC FeSe on SrTiO$_3$.} \textbf{a}, Plot of the Fermi surface with only electron pockets located at the zone corner (M-point). Red dots are the approximate points on the Fermi surface where gaps were extracted in Fig. 2e-f. \textbf{b}-\textbf{c}, High-symmetry cuts along the purple line plotted in \textbf{a}, taken at 16 K. \textbf{b} and \textbf{c} are centered at $\Gamma$ (zone center) and M, respectively. All such cuts in this paper are given in units of inverse angstroms relative to the high symmetry points. The hole band seen in \textbf{b} is located 80 meV below E$_{F}$. In \textbf{c} a different color-scale highlights two important features: the electron band with a minimum at 60 meV below E$_{F}$ (labelled \textbf{A}), and a replica electron band (labelled \textbf{A$^{\prime}$}), which is located 100 meV below the former and sits on top of a broad hole band. \textbf{d}, Energy distribution curves (EDCs) at M shown as a waterfall plot, with markers indicating band peaks. \textbf{e}-\textbf{f}, Second derivatives in energy of the high symmetry cuts from \textbf{b} and \textbf{c}. An additional weaker replica, labeled \textbf{C}, can now be seen at M in \textbf{f}, sitting below \textbf{A}, and at the $\Gamma$ point in \textbf{e} we see the hole band and a corresponding replica, labelled \textbf{D} and \textbf{D$^{\prime}$}, respectively. \label{fig1}}
\end{figure}

\begin{figure}
\includegraphics[scale=1]{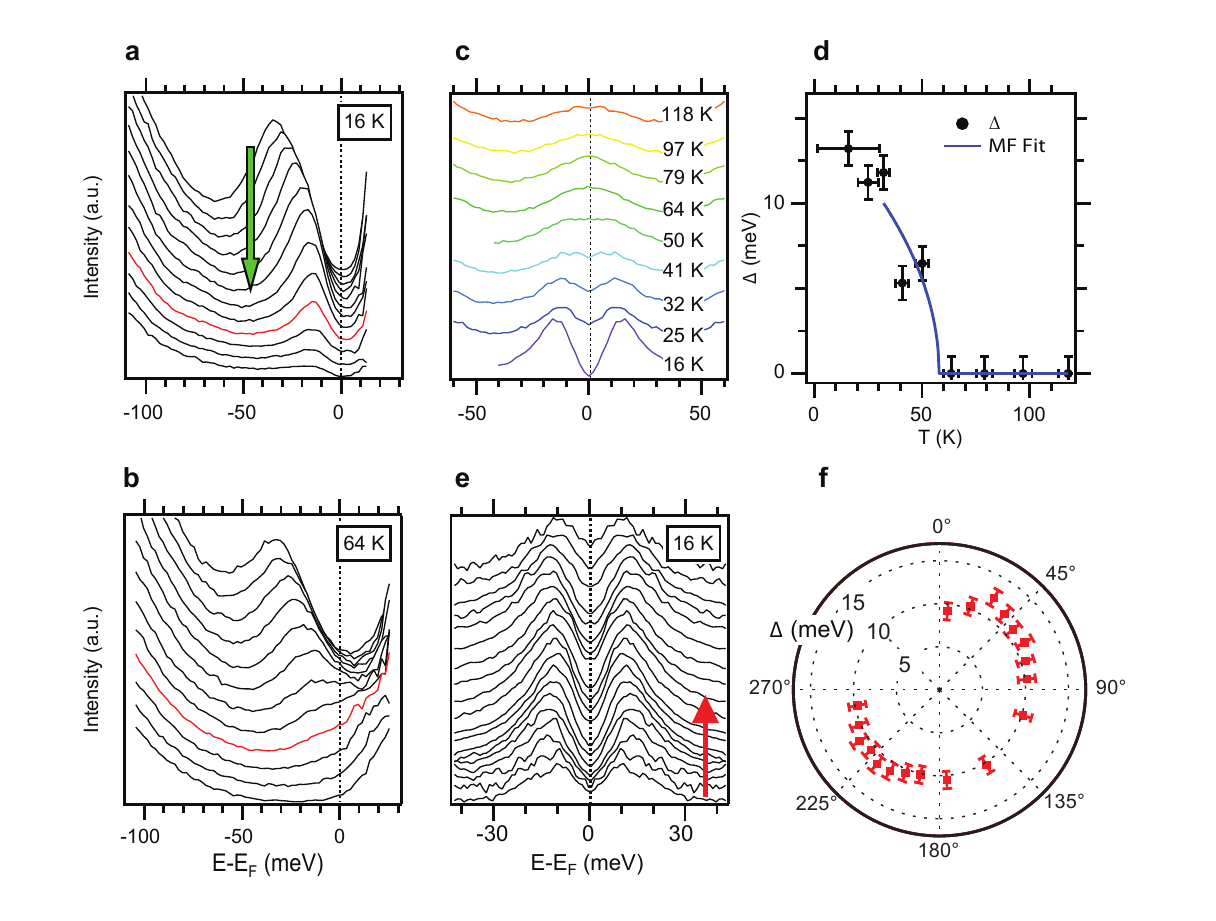}
\caption{\textbf{Temperature dependence of the 1UC film energy gap at the M-point.} (\textbf{a}-\textbf{b}), Plots of the $\Gamma$-M high symmetry dispersion EDCs at 16 K and 64 K respectively, with the Fermi-Dirac distribution factored out. The cut direction is given by the green arrow marked in Fig. 1c The red trace represents the EDC at the Fermi momentum ($k_{F}$). \textbf{c}, Plot of the evolution of the symmetrized EDCs at $k_{F}$ as a function of temperature, where we observe a gap closing between 50 K and 64 K. \textbf{d}, The gap evolution as a function of temperature, with gap extracted using the model in Ref.\citen{PhysRevB.57.R11093}. Error bars include drift of E$_{F}$ as measured relative to a gold reference. A fit to a mean-field type of order parameter is plotted, giving a gap closing temperature of 58 K. \textbf{e}, Symmetrized EDCs at different $k_{F}$'s along the Fermi surface at 16 K, in the direction indicated by the red arrow in Fig. 1a. \textbf{f}, Polar plot of the gap of the EDCs from \textbf{e}.}
\end{figure}

\begin{figure}
\includegraphics[scale=1]{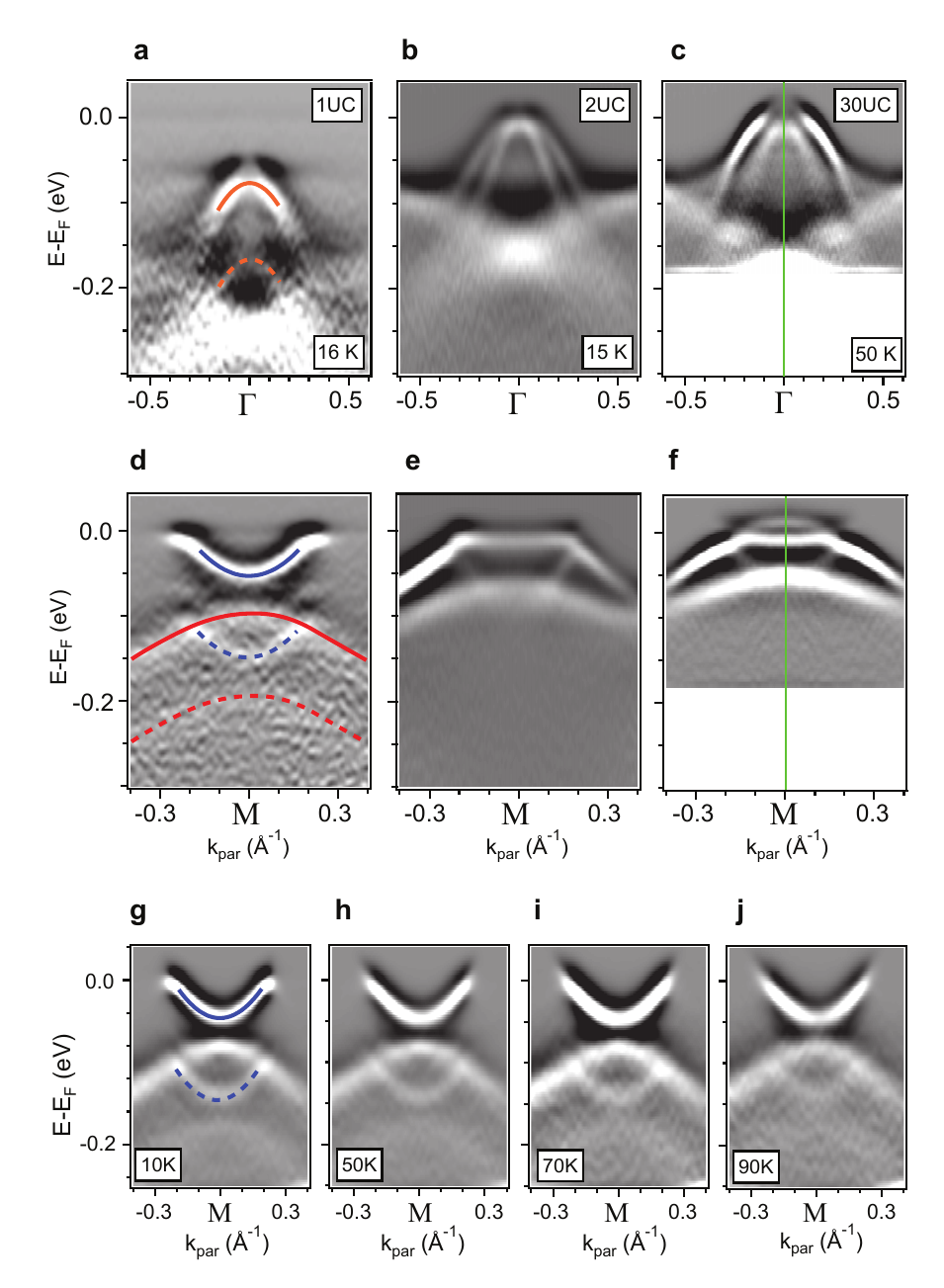}
\caption{\textbf{Dependence of electronic structure on FeSe film thickness.} \textbf{a}, Spectrum of the 1UC film at $\Gamma$. \textbf{b}, $\Gamma$ spectrum for the 2UC film. \textbf{c}, $\Gamma$ spectrum for the 30UC film.  \textbf{d}, Spectrum of the 1UC film at M. \textbf{e}, M spectrum for the 2UC film. \textbf{f}, M spectrum for the 30UC film. Data in \textbf{c} and \textbf{f} have been symmetrized around $\Gamma$ and M, respectively, as indicated by the green line, while data on the 1UC and 2UC films are unsymmetrized. The colored lines in \textbf{a} and \textbf{d} are guides to the eye, with solid lines denoting the main band and dashed lines corresponding to the replicas. The main bands and replicas are color-coded according to Fig. 1. We do not observe replica bands for either the 2UC and 30UC films. Raw spectra can be found in Extended Data Fig. 2. \textbf{g}-\textbf{j}, Temperature dependence of the replica bands, which persist at temperatures greater than the gap-opening temperature. \label{fig3}}
\end{figure}

\begin{figure}
\includegraphics[scale=1]{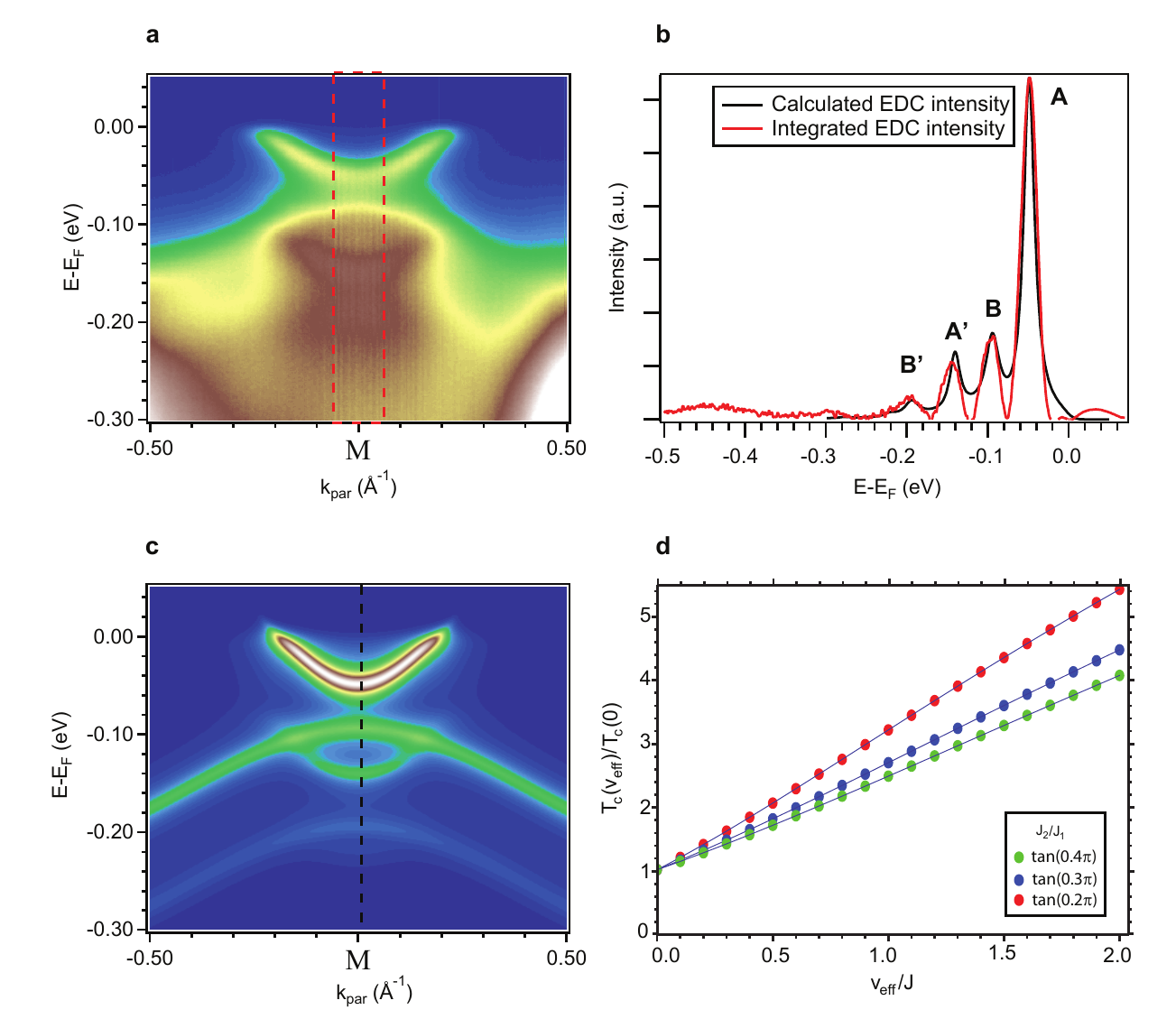}
\caption{\textbf{Extraction of the electron-phonon coupling and determination of T$_c$ enhancement.} \textbf{a}, High statistics scan at M taken at low temperature (10 K). The spectral weight is integrated over the momentum range indicated by the dotted rectangle to obtain better statistics for a single EDC. \textbf{b}, The integrated EDC at M (after background subtraction) and the EDC from our calculations. Peaks corresponding to the bands are labeled according to Fig. 1a. \textbf{c}, Model spectral function calculation including both hole and electron bands coupled to a dispersionless 80 meV phonon mode (see SI for details). The black dotted line indicates the EDC plotted in \textbf{b}. \textbf{d}, Plot of the T$_c$ enhancement as a function of effective attractive electron-electron interaction strength ($v_{\rm eff}/J$). Shown are plots for three different values of $J_2/J_1$. The parameters we used to construct this curve (see SI for additional details) are $q_0=0.1\pi/a$ ($a = 3.9$ \AA), $J=$ 30 meV, cutoff energy = 65 meV, and $T_c$ = 40 K in the absence of electron phonon interaction. With the extracted parameters, we place the lower bound of the enhancement factor at 1.5.\label{fig4}}
\end{figure}

\begin{EDfigure}
\includegraphics{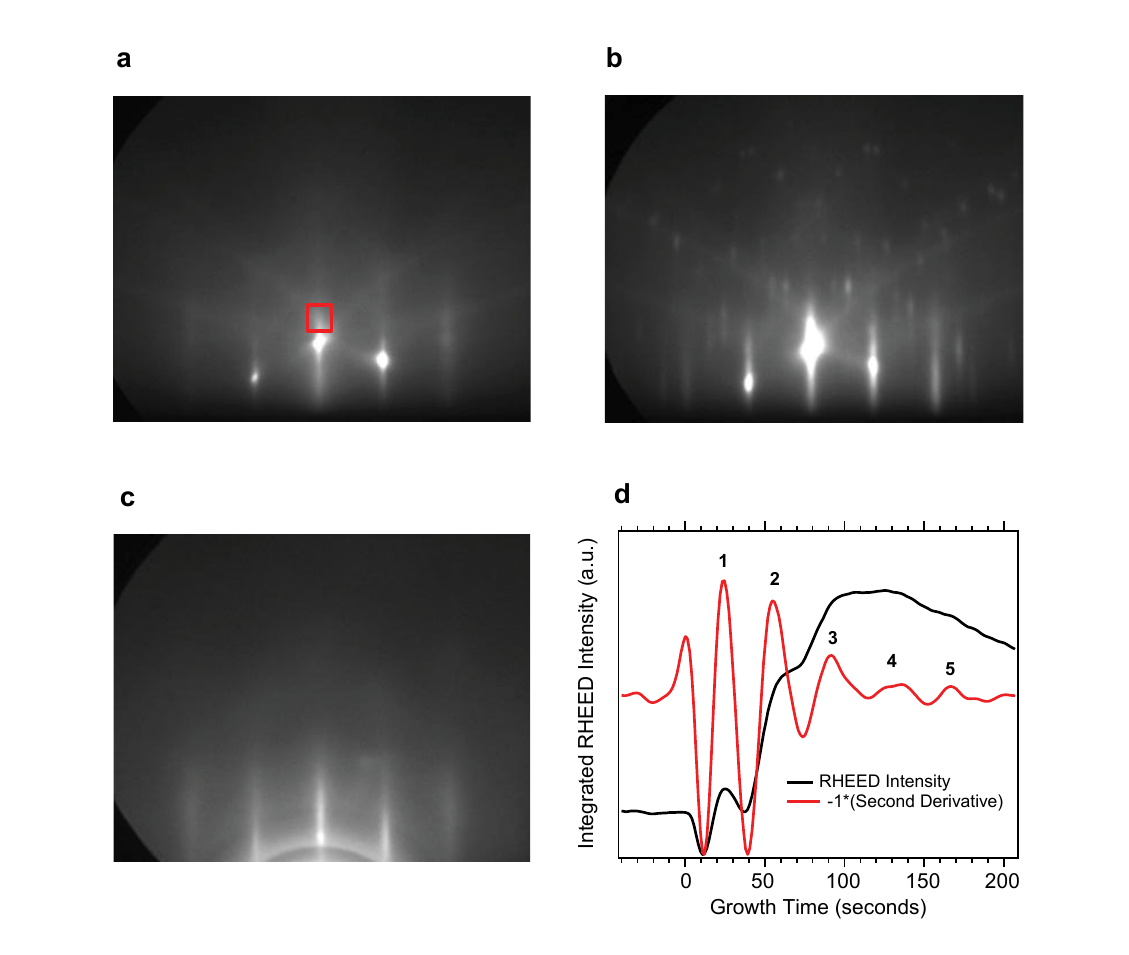}
\caption{\textbf{RHEED images observed during FeSe growth.}  \textbf{a}, RHEED image of
STO substrate after degassing at 450 $^\circ$C for 1 hour. Red box highlights
the region integrated for monitoring RHEED oscillations.  \textbf{b}, Surface
reconstruction as observed by RHEED at annealing temperatures.  \textbf{c}, RHEED image
of FeSe 1UC film showing uniform streaks typical of an atomically flat
thin film. \textbf{d}, RHEED intensity for integration region shown in \textbf{a} (black).
The second derivative of the intensity curve (red) highlights the RHEED
oscillations signaling the completion of a unit cell after $\approx 30$
seconds.} 
\label{RHEED}
\end{EDfigure}

\begin{EDfigure}
\includegraphics[scale=1]{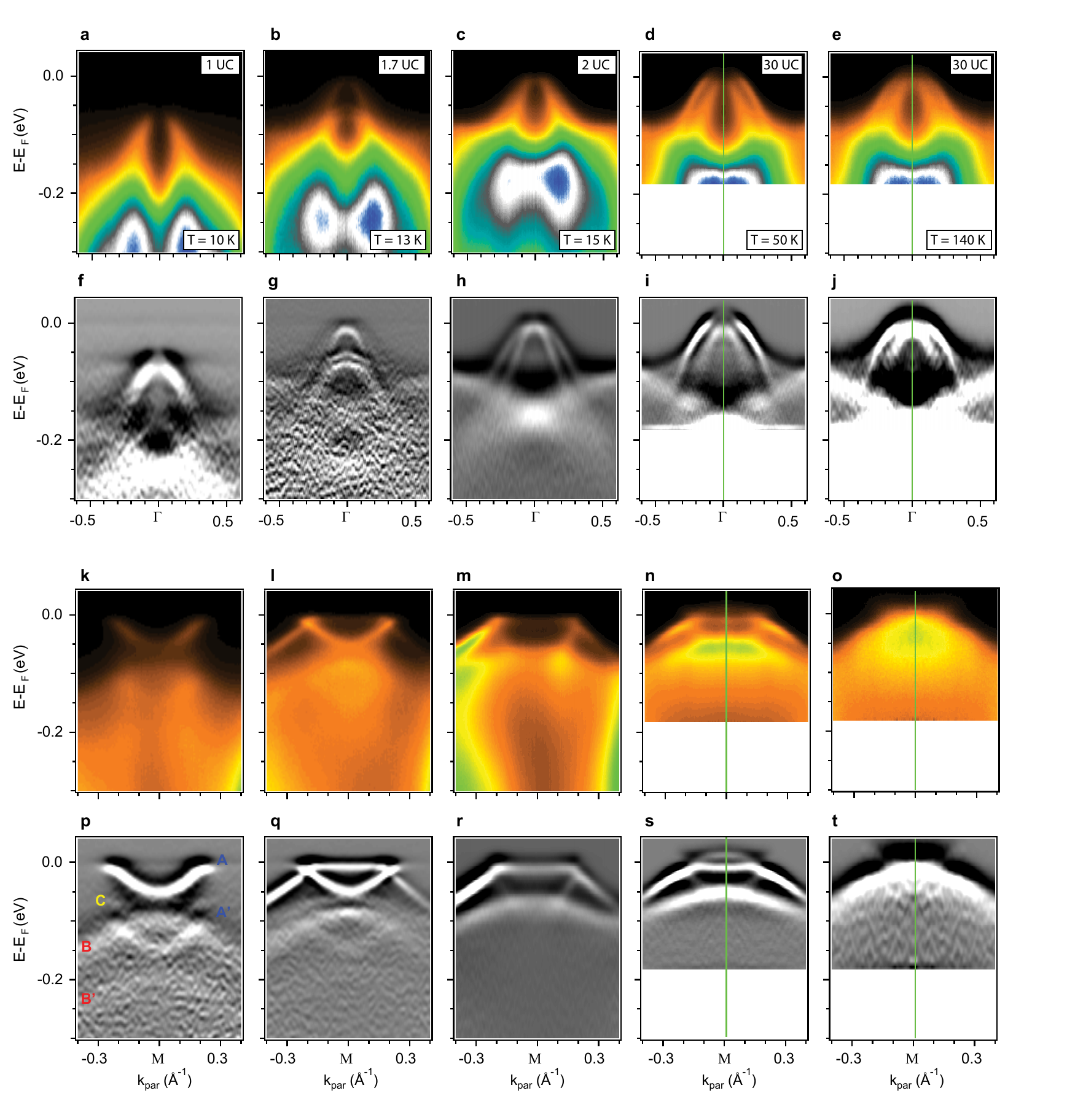}
\caption{\textbf{Raw spectra and second derivatives of 1UC, 1.7UC, 2UC, and 30UC films.}
\textbf{a}, \textbf{f}, \textbf{k}, \textbf{p}, Plots of the raw spectrum at $\Gamma$, the second derivative at $\Gamma$, the raw spectrum at M, and the second derivative at M for the 1UC film, respectively. \textbf{b}, \textbf{g}, \textbf{l}, \textbf{q}, Plots of the raw spectrum at $\Gamma$, the second derivative at $\Gamma$, the raw spectrum at M, and the second derivative at M for the 1.7UC film, respectively. \textbf{c}, \textbf{h}, \textbf{m}, \textbf{r}, Plots of the corresponding data for the 2UC film. \textbf{d}, \textbf{i}, \textbf{n}, \textbf{s}, Plots for the 30UC film at 50 K. \textbf{e}, \textbf{j}, \textbf{o}, \textbf{t}, Plots for the 30UC film at 140 K. Data for the 30UC film are symmetrized about the high symmetry points (indicated by the green line), and were taken with 25 eV photons. We observe a band splitting in the 30UC film at M (\textbf{s}), at low temperature. This band splitting closes at higher temperature (140 K), where we now observe only one band (\textbf{t}).}
\end{EDfigure}

\begin{EDfigure}
\includegraphics[scale=1]{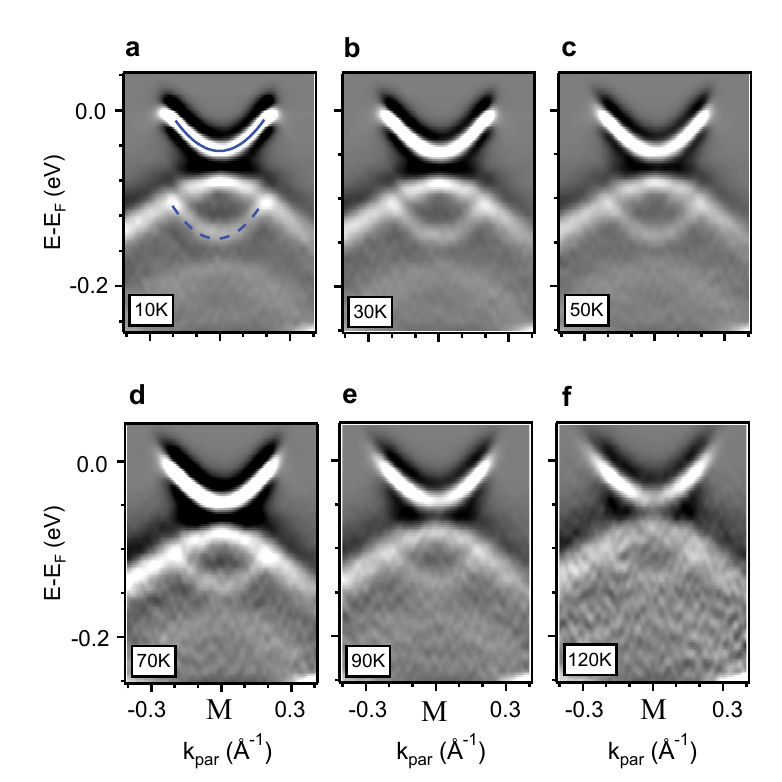}
\caption{\textbf{Temperature evolution of the M point spectrum of the 1UC film.} \textbf{a}, Spectrum at 10 K, where we can clearly see the bend-back of the replica bands, exactly like the main bands near E$_F$. \textbf{b}, Spectrum at 30 K. \textbf{c}, Spectrum at 50 K. \textbf{d}, Spectrum at 70 K. \textbf{e}, Spectrum at 90 K. \textbf{f}, Spectrum at 120 K. The replica bands persist up to temperatures significantly higher than the gap-opening temperature. }
\label{Tdep}
\end{EDfigure}

\begin{EDfigure}
\includegraphics[scale = 1] {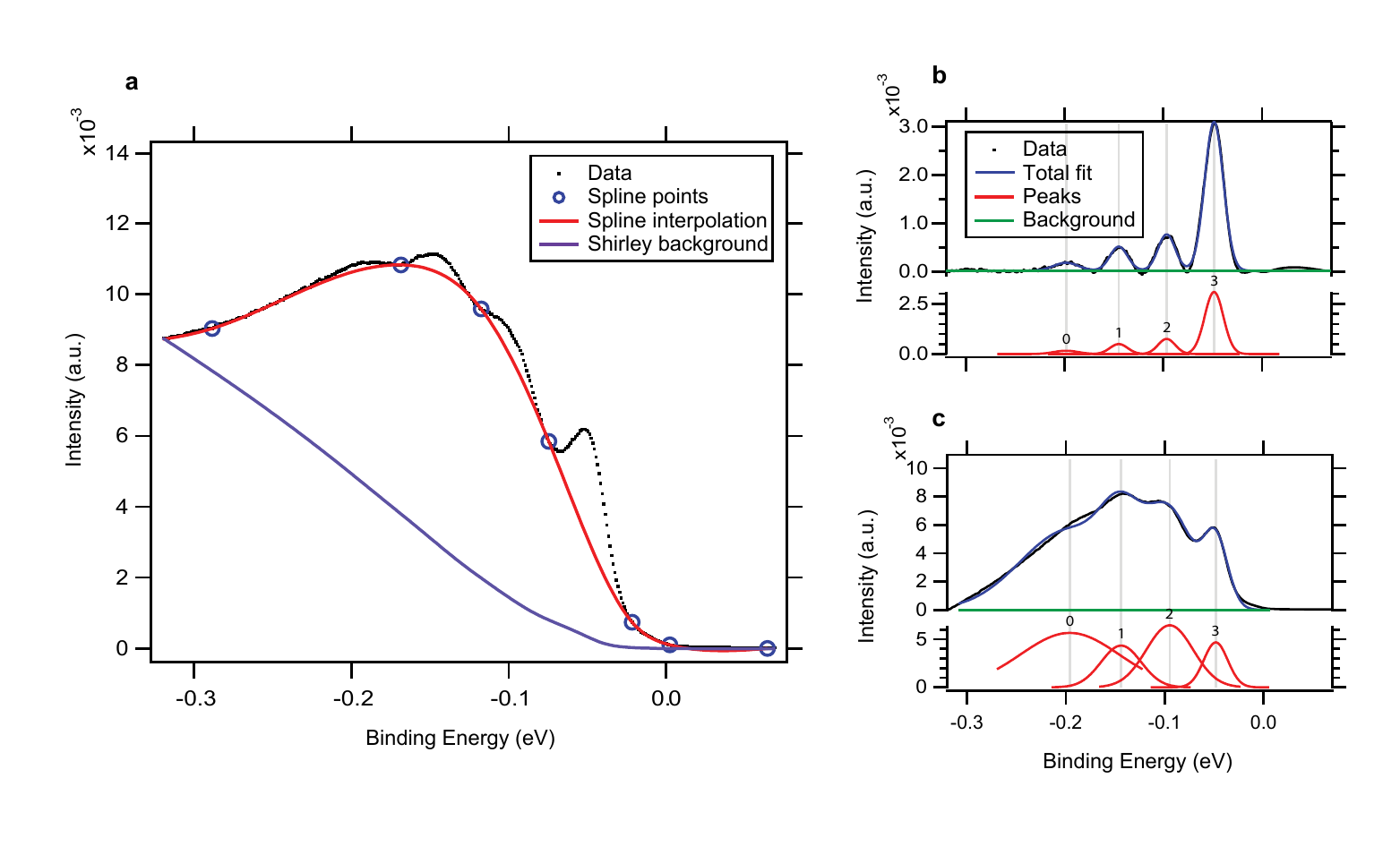}
\caption{\textbf{Fitting the intensities of the ARPES spectra at M.} \textbf{a}, Plot showing two
different backgrounds used in the fitting.  Using the blue circles as fixed
points, we first modeled the background using a spline interpolation, plotted
in red. The second fit used a Shirley background (see SI), plotted in purple. \textbf{b}, Data and fitting with the spline background subtracted. We fit to four Gaussian peaks, which are plotted separately for clarity. \textbf{c}, Data and fitting with the Shirley background subtracted. We restrict our fitting energy window from -0.32 eV to 0.03 eV.}
\label{fitting}
\end{EDfigure}

\begin{EDfigure}
\includegraphics[scale = 1]{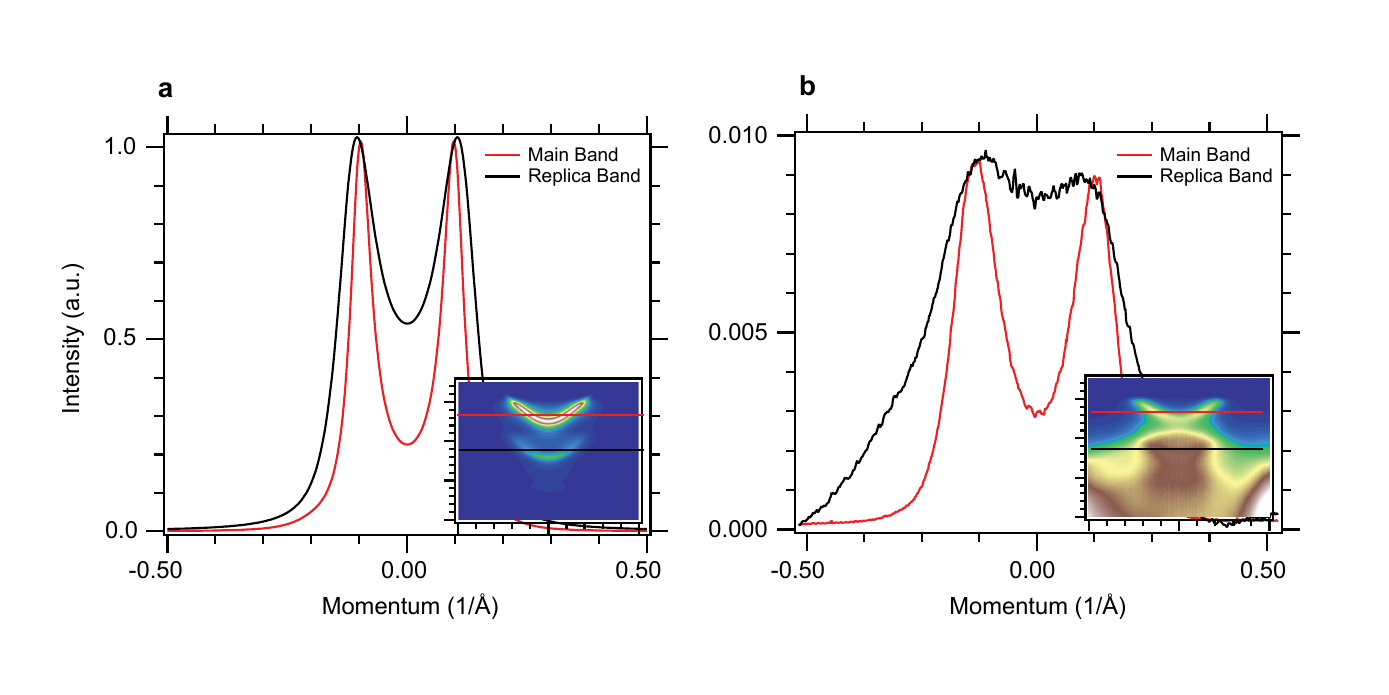}
\caption{\textbf{Momentum distribution comparison between the main band and replica band.} \textbf{a}, The momentum distribution curves (MDCs) of our theoretical calculation of the main electron band and the replica electron band with normalized intensities. The MDCs of both bands are taken at the same energies with respect to their band bottoms (see inset). The replica band peaks are broadened due to the elctron-phonon coupling. \textbf{b}, The MDCs of our data, with a momentum  independent background subtracted from the replica band MDC. The momentum-dependent background, e.g contributions from the hole band, is the likely cause of the extra broadening in the data. }
\label{mdcs}
\end{EDfigure}

\begin{EDfigure}
 \begin{centering}
 \includegraphics[width=0.8\textwidth]{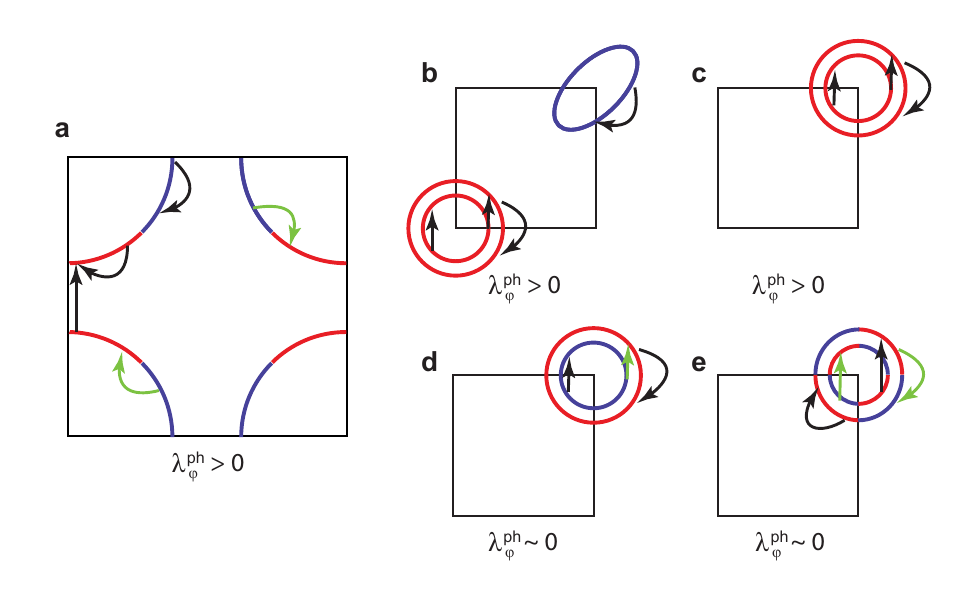}
 \caption{\label{Fig:GapSketch}\textbf{Effects of electron-phonon coupling on different gap symmetries and Fermi surfaces.}
  Cartoon sketches of the various Fermi surfaces and gap symmetries found or proposed in unconventional superconductors. \textbf{a}, Sketch for the cuprates with a $d$-wave gap. \textbf{b-e}, Various scenarios for the iron-based superconductors. Only one quarter of the first Brillouin zone is shown for clarity. The thick blue and red lines indicate the phase of the gap. The arrows show various forward-focused scattering processes. The black arrows indicate scattering processes that connect portions of the Fermi surface with the same sign gap and are therefore pair enhancing. The green arrows show pair-breaking processes which connect regions of the Fermi surface with different signs.}
 \end{centering}
\end{EDfigure}

\begin{EDfigure}
\includegraphics[scale=1]{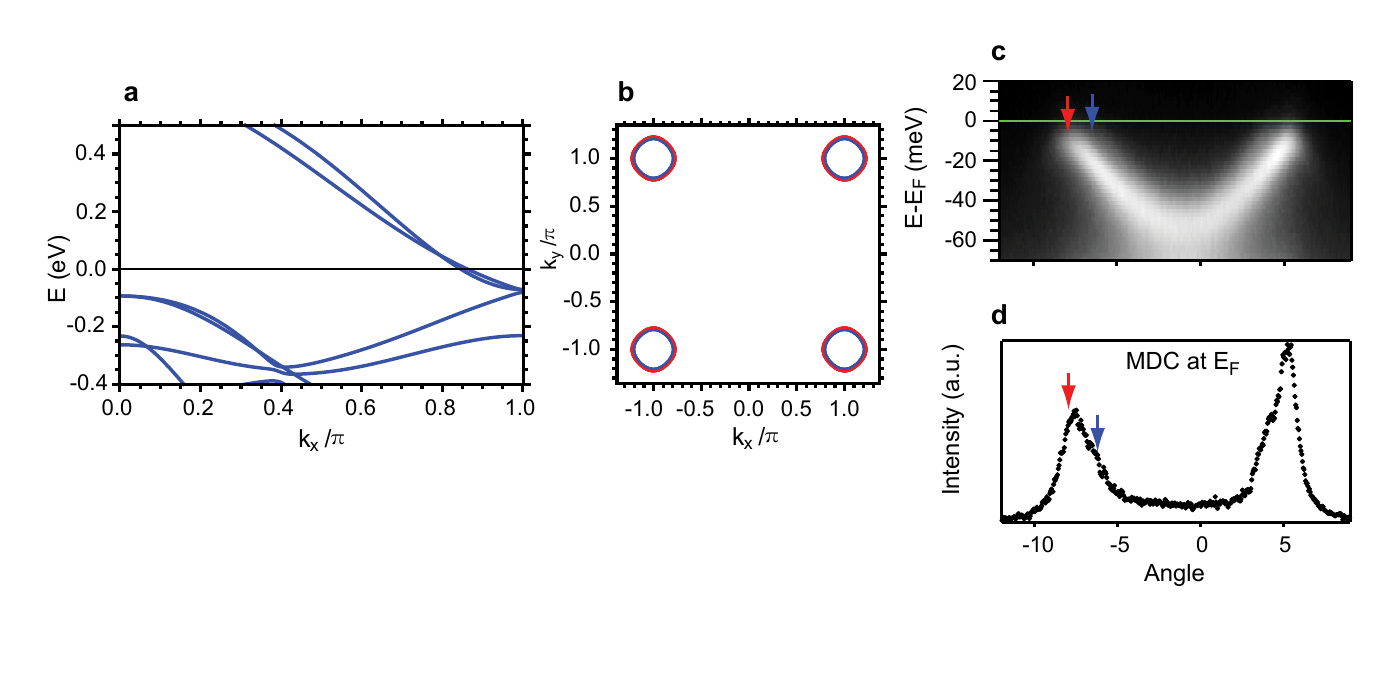}
\caption{\textbf{Input electronic structure for calculated T$_c$ enhancement.} \textbf{a}, Calculated band structure used in our determination of T$_c$ enhancement. \textbf{b}, Calculated Fermi surface showing slightly split electrons pockets. \textbf{c}, Dispersion along M-point showing two nearly degenerate bands. \textbf{d}, Momentum distribution curve (MDC) at E$_F$ showing peaks from two bands.}
\label{band}
\end{EDfigure}

\begin{EDfigure}
\includegraphics[scale=1]{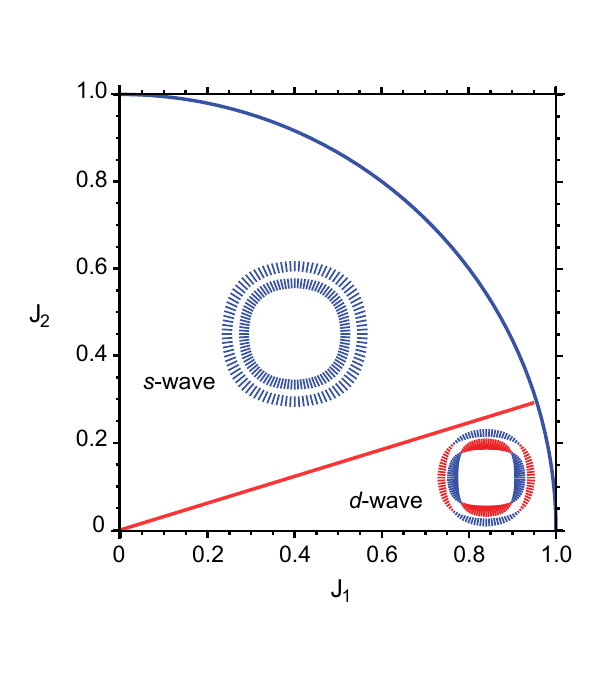}
\caption{\textbf{Phase diagram of the $J_1-J_2$ model.} The blue line represents $\sqrt{J_1^2 + J_2^2}=1$. The red line represents the transition between different gap symmetries at $J_2/J_1$$\sim.31$. Above the transition one finds a gap with $s$-wave symmetry. Below the transition one finds a gap with $d$-wave symmetry.  Diagrams of the two different possible symmetries are drawn in their respective region of the phase diagram. The length of the tick lines represent the magnitude of the gap, while the color represents the sign (red: minus, blue: plus). The two electron pockets in the figure are separated for clarity.}
\label{GapJ2J1}
\end{EDfigure}



\end{document}